\documentclass{aa}  
\usepackage{graphicx}
\usepackage{txfonts}
\usepackage{natbib,twoopt}
\bibpunct{(}{)}{;}{a}{}{,}             
\usepackage{hyperref}
\hypersetup{
    colorlinks=true,
    linkcolor=blue,
    citecolor = blue,
    filecolor=magenta,
    urlcolor=blue}
\usepackage{booktabs}

\begin{document}

\authorrunning{Origlia et al.}
\titlerunning{The link between Terzan~5 and the Galactic bulge}

\title{On the manifest link between Terzan~5 and the Galactic bulge}

\author{L. Origlia\inst{1}, F. R. Ferraro\inst{2,1},
    C. Fanelli\inst{1},
    B. Lanzoni\inst{2,1},
    D. Massari\inst{1},
    E. Dalessandro\inst{1} and C. Pallanca\inst{2,1}
    }

   \institute{INAF, Osservatorio di Astrofisica e Scienza dello Spazio di Bologna, Via Gobetti 93/3, I-40129 Bologna, Italy
    \email{livia.origlia@inaf.it}
         \and
         Dipartimento di Fisica e Astronomia, Università degli Studi di Bologna, Via Gobetti 93/2, I-40129 Bologna, Italy
         }

\abstract{We address the chemical link between Terzan~5 (hereafter Ter5) and the Bulge, as probed by the observed distributions of [$\alpha$/Fe] abundance ratios with varying [Fe/H] and by suitable statistical tests to evaluate their significance. We also present a comprehensive review of the kinematic and evolutionary properties of Ter5, based on all the available observational signatures and  the scenarios proposed so far in the literature for the formation and evolution of Ter5,  based on these observational facts and the recent modeling of its star formation and chemical enrichment history. 
This analysis confirms the complex nature of this massive stellar system, with robust evidences of a bulge in-situ formation and of a subsequent evolution that cannot be simply explained by a single merging/accretion event of two globulars or a globular and a giant molecular cloud, as proposed in the literature, but it requires a more complex star formation likely accompanied by some self-enrichment.}


\maketitle

\section{Introduction}
The formation and early evolution of the Galactic bulge and more generally, of galaxy bulges is a topic largely debated in the literature.
Early (gas and stars) merging, friction of massive clumps and proto-disk evaporation \citep[e.g.,][]{imm04, dek09}, and/or dynamical interactions with the central disk/bar \citep[e.g.,][]{ger12,saha13} have been proposed in the literature.
High-redshift observations of the so-called “clumpy or chain galaxies” \citep[e.g.,][]{elm09,gen11,tac15} show the existence of massive clumps of gas and stars that plausibly provide the “set of initial conditions” for the assembling process of galactic spheroids. In fact, numerical simulations \citep[e.g.,][]{imm04,elm08,bournaud09,bournaud16} show that such massive clumps (with masses of 10$^8$-10$^9$ M$_{\odot}$) can form from violent disk instabilities in gas-rich galaxies, and/or by the clustering of smaller, seed clumps of  10$^7$-10$^8$ M$_{\odot}$ \citep[e.g.,][]{beh16} in a bottom-up scenario. They eventually migrate to the centre and coalesce in a dissipative way to generate the bulge. These models naturally provide short timescales (a few 10$^8$ yr) for the bulge assembly, with intense star formation and chemical enrichment by type II supernova (SN) ejecta, leading to a metal-rich and $\alpha$-enhanced stellar population, in agreement with the chemical abundances measured in the Galactic bulge \citep[e.g.][]{johnson_14}. In such scenario, it is possible that some fragments from the pristine massive clumps survive the total disruption and evolve as independent stellar systems with the appearance of the massive globular clusters (GCs) in the inner regions of the host galaxy. At odds with genuine GCs, however, these fossil relics  
are expected to host multi-iron and multi-age sub-populations, because their progenitors could have been massive enough to retain the iron-enriched ejecta of SN explosions and likely experienced multiple bursts of star formation. Clearly, finding stellar systems with the predicted properties in the Galactic bulge, not only would be of paramount importance for our understanding of the Milky Way (MW) history, but it would also provide a unique window on the formation of spheroids in general. Of course, the best systems to investigate for this purpose are bulge GCs. 

Bulge GCs are remarkable in being massive and old, yet having survived in a dynamically harsh environment. Among all Galactic GCs, they are the least studied so far, especially because of the prohibitive conditions of high and differential reddening, combined with high stellar crowding, that are found along their line of sight. Over the last years, we have been engaged in a long term project aimed at characterizing the structure, age, kinematics and detailed chemical composition of bulge GCs using VLT, Keck, Gemini and HST instrument facilities \citep{origlia02, origlia04, origlia05, origlia08, valenti07, valenti10, valenti11, valenti15, saracino15, saracino16, saracino19, pallanca19, pallanca21a, pallanca23, cadelano20, dalessandro22, cadelano23, deras23, leanza23}.

The most striking result was the discovery that Ter5, commonly catalogued as a GC of the inner bulge and affected by large and differential reddening \citep[][]{valenti07,massari12}, is actually a complex stellar system with sub-populations of significantly different metallicities and ages. In particular, adaptive optics imaging with VLT-MAD and some spectroscopy with Keck-NIRSPEC in the near IR \citep{ferraro09}, revealed two distinct red clumps in the color-magnitude diagram with very different iron abundances ([Fe/H]=-0.2 and +0.3 dex, respectively) and likely with significantly different ages, later confirmed  by the detection of two distinct main-sequence turnoff points, providing ages of 12 Gyr for the (dominant) sub-solar component and 4.5 Gyr for the super-solar one \citep{ferraro16}. 
The comprehension of the true nature of Ter5 has become even more intriguing after the discovery that a second massive Bulge stellar system, namely Liller~1, hosts two main distinct subpopulations \citep{ferraro21} with remarkably different ages and metalllicities: a $\sim 12$ Gyr old subpopulation with sub-solar metallicity has been found to cohabit with at least one additional component showing an age of just 1-2 Gyr and a super-solar iron abundance; as in Ter5 \citep[see][]{lanzoni10}, the youngest and most metal-rich sub-population is more centrally concentrated than the other(s).
These peculiar properties unambiguously probe that Ter5 and Liller~1 are not genuine GCs, and call for more complex formation scenarios.  

In this paper we first critically review the kinematic and chemical properties of the Ter5 stellar subpopulations
from the studies published so far in the literature after the 
\citet{ferraro09} discovery of its complex nature, and the proposed scenarios for its formation and evolution.
Then we focus the discussion on the chemical link between Ter5 and the bulge since of crucial importance to constrain the in-situ formation hypothesis, by presenting a new, comprehensive statistical analysis of suitable chemical abundance pattern distributions from different high resolution spectroscopic surveys, including The Apache Point Observatory Galactic Evolution Experiment (APOGEE, \citealt{apogee17}). Finally, we summarize the current view of Ter5, also based on recent modeling of its star formation and chemical enrichment history, and we briefly address the remaining open questions and future perspectives.

\section{Properties of Ter5 stellar subpopulations}
\label{properties}
The chemistry of Ter5 stars has been characterized in some detail through a rather massive spectroscopic follow-up in both the NIR and the optical bands. From the analysis of high-resolution Keck-NIRSPEC spectra in the H-band we measured the abundances of Fe, C, O and a few other $\alpha$ and light elements for almost 40 giant stars \citep{origlia04, origlia11, origlia13}. From VLT-FLAMES and Keck-DEIMOS spectra in the Ca-T region, we obtained Fe abundances for more than a hundred giants likely members of the system \citep{massari14a}, and an other hundred in a surrounding control field \citep{massari14b}. 
The membership to Ter5 of these stars has been established according to their radial velocity (RV) and for a sub-sample of them it has been also confirmed by proper motion measurements \citep{massari15}.
The resulting metallicity distribution of Ter5 has revealed the presence of two main peaks at sub-solar and super-solar iron abundances ([Fe/H]$\sim -0.3$ dex and $\sim +0.3$ dex, respectively), and an additional, minor component at [Fe/H]$\approx-0.8$ dex, bringing the overall metallicity range covered by the system to approximately 1 dex. 
The metal-poor subpopulations of Ter5 are $\alpha$-enhanced ([$\alpha$/Fe]$\sim +0.3$ dex) and likely formed early and quickly from a gas evidently polluted by a large number of type II SNe. The young, metal-rich subpopulation, more centrally concentrated, has an approximately solar-scaled [$\alpha$/Fe] abundance ratio, requiring a progenitor gas further polluted by both type II and Ia SNe on a longer timescale. 
Although \citet{pfe21} suggested that the metal poorest stars (with [Fe/H]$\simeq -0.8$ dex) may have been accreted from the surrounding, \citet{massari14b} have shown that this component 1) is not negligible (indeed, after statistical decontamination from the Galactic contribution, it amounts to $\approx 6\%$ of the whole Ter5 population, while field stars are significantly more marginal: $\approx$3\%; see also \citealt{massari14a}), and 2) it appears to be similarly centrally concentrated as the main subpopulation, thus making somewhat unlikely the accretion hypothesis.
It is also worth mentioning that the membership of these metal-poor stars to Ter5 has been established from both their RVs and their HST proper motions \citep{massari15}. 

More recently, we also measured the chemical abundances of nine variable stars (three newly discovered RR Lyrae and six known Miras) from VLT-XSHOOTER NIR spectra \citep{origlia19}. 
The three RR Lyrae have radial velocities and proper motions \citep{massari15} fully consistent with being Ter5 members. Their [Fe/H]$\approx-0.7$ dex and [$\alpha$/Fe] enhancement nicely matches the chemistry of the most metal-poor subpopulation of the system, thus directly probing an old age for this component. The three short-period (P$<$300 days) Miras are also consistent with being members of Ter5 based on their radial velocities and they show [Fe/H]$\approx -0.3$ dex, enhanced [$\alpha$/Fe], and a mass consistent with old ages, thus well corresponding to the main, 12 Gyr-old and sub-solar component of Ter5. Only one, out of the three long-period (P$>$300 days) Miras turned out to likely belong to the system, and it displays a super-solar iron abundance, solar-scaled [$\alpha$/Fe] and a mass consistent with being several Gyr younger than the old, short-period Miras, thus nicely matching the properties of the young, metal-rich subpopulation.  
Based on these results, the metallicity distribution of Ter5 closely resembles the one of the surrounding bulge field \citep{massari14a, massari14b}, although it is somehow more peaked around the two main subpopulations. The [$\alpha$/Fe] {\it vs} [Fe/H] distribution drawn by the different components of Ter5 \citep{origlia11, origlia13}, with the typical knee occurring at almost  solar metallicity, thus implying a very high star formation rate, also fully matches the corresponding bulge distribution \citep[e.g.,][and references therein]{johnson_14}, while it is at odd with those of the other MW substructures (i.e. thin and thick discs and halo) or with those of satellite galaxies in the Local Group\citep[see e.g.][]{matteucci90,tolstoy09}. 

Among the formation scenarios proposed so far in the literature for Ter5\footnote{While here we discuss specifically the case of Ter5, analogous arguments hold for Liller 1 as well. The most relevant information for Liller~1 can be found as follows: the reconstructed orbit in \citet{baum19}, the observed metallicity distribution in \citet{crociati_23}, detailed chemical abundances for $\alpha$-elements and other metals in \citet{deimer24, fanelli24}, the reconstructed star formation history in \citet{dalessandro22}.}, the possibility that it could be the former nuclear star cluster of an accreted dwarf galaxy \citep{bro18,alf19} can be discarded on the basis of several arguments.
First, the age-metallicity relation of Ter5 turns out to be consistent with the MW enrichment \citep[e.g.,][]{sna15,kru19}, thus providing evidence of a Galactic origin of the system. 
Second, the reconstruction of the Ter5 orbit from the measured radial velocities and proper motions  \citep{massari15, massari19, vasiliev_21} leads to an apocenter of just 2.8 kpc, which confines the system well within the bulge and therefore points to an \emph{in-situ} formation.
In addition, as also discussed in \citet{pfe21}, the accretion scenario would require 1) a host galaxy with a mass similar to the MW one, in order to have a nuclear star cluster with a metallicity consistent with that of Ter5 and of the MW nuclear star cluster \citep{rich17,fel17}, and 2) a major merger occurred within approximately the last 4 Gyr (i.e., more recent than the age of the youngest sub-population in Ter5), for which there is no evidence in the MW \citep[indeed, our galaxy seems to have escaped any significant merger over the last 10 Gyr; see, e.g.,][and references therein]{ham07,helmi18,fra20}.

Consistently with an in-situ origin and evolution of Ter5, the possibilities of the merger between two GCs \citep{khoperskov18, mastrobuono19, pfeffer21, ishchenko23}, or the accretion of a giant molecular cloud by a genuine GC \citep{mckenzie18, bastian22} have been proposed in the literature. 
However, both the old and sub-solar component, and the younger and super-solar population of Ter5 are very massive, reaching about $10^6$ M$_{\odot}$ \citep{lanzoni10}, thus exceeding the mass of the largest Galactic GCs, like 47 Tucanae \citep{harris96} and NGC 6388 \citep{lanzoni07}. Moreover, the super-solar component is significantly younger and more metal-rich than any Galactic globular and largely more massive than any Galactic open cluster.
Finally, the observed metallicity distribution \citep[e.g.][]{massari14b} and the reconstructed star formation history of the system \citep{crociati24} indicate the presence of at least three sub-populations and/or a prolonged star formation, which cannot be accounted by a single merging event.

\begin{table}
\scriptsize
\renewcommand{\arraystretch}{1.25}
\setlength{\tabcolsep}{5.5pt}
\caption{Scaling factors to the \citet{magg_22} solar reference abundances.}
\label{tab1}
\begin{tabular}{|c|c|c|c|c|c|}
\hline\hline
Dataset   & $\Delta$[Fe/H]   & $\Delta$[O/H]   & $\Delta$[Mg/H]   & $\Delta$[Si/H]   & $\Delta$[Ca/H]   \\
\hline
 [Johnson14] - [Magg22]     & 0.02  &  -0.08  &  +0.03  &  -0.04  &  -0.01  \\
 \hline
 [Origlia,Rich] - [Magg22]  & 0.00  &  +0.06  &  +0.03   &  +0.04  &  +0.01  \\
 \hline
 [Bensby14]- [Magg22]  & 0.00  &  -0.08  & +0.05   &  -0.08  &  -0.03  \\
\hline\hline
\end{tabular}
Note: Solar reference abundances adopted by \citet{johnson_14} are reported in their Table~2, those 
adopted by \citet{rich05,rich07,rich12,origlia04,origlia11,origlia13} are from \citet{Grevesse98}, and  those adopted by \citet{bensby14} are from \citealt{Asplund09}.
\end{table}

In this respect, chemical models specifically computed for Ter5 \citep{romano23} have demonstrated that the self-enrichment
of a few $10^7$~M$_{\odot}$ progenitor can nicely reproduce all the observed chemical patterns. 
Other intriguing characteristics of Ter5 are its large population of millisecond pulsars (MSPs), that   accounts for about 25\% of the entire MSP population found so far in Galactic GCs \citep{ransom05,cadelano18,pad24}, and its surprising large collision rate (the largest among all Galactic GCs  \citealt{verbunt87,lanzoni10}).

The picture proposed by our group \citep{ferraro16} is that Ter5 (like Liller 1; \citealp{ferraro21,dalessandro22,fanelli24}) may be the relic of a pristine massive (a few $\sim 10^7 M_\odot$) clump of gas and stars of the early bulge that survived total disruption and evolved independently, likely experiencing multiple episodes of star formation and self-enrichment. This is consistent with the evidence of an in-situ origin, the large estimated masses of its subpopulations, and the highest central concentration of the metal-richer component. It can also accommodate the presence of multiple subpopulations and the Ter5 complex chemical evolution \citep{romano23} and star formation history.

\section{The chemical link between Ter5 and the bulge}
\label{link}

The chemical, kinematic and evolutionary properties of Ter5 discussed in the literature and summarized in Sect.~\ref{properties} are consistently supporting an in-situ bulge formation and evolution of Ter5. However, \citet{taylor22} questioned such an evidence by means 
of a chemical study based on APOGEE spectra of 5-7 giants candidate members of Ter5, whose abundance patterns are claimed to show some deviation from the bulge distribution.
In order to shed new light on this issue and to characterize the chemical properties of Ter5 stellar subpopulations and their possible link with the bulge on a more quantitative ground, we performed a few statistical tests on the Ter5 and bulge distributions of suitable element abundances and abundance ratios, 
namely [Si/Fe], [Ca/Fe], [Mg/Fe], [O/Fe] and the average [<Ca,Si,Mg,O>] (hereafter, [$\alpha$/Fe]) trends with [Fe/H]. [$\alpha$/Fe] {\it vs} [Fe/H] distributions are indeed primary indicators of the chemical enrichment history of a given stellar system, being $\alpha$-elements and iron synthesized in stars with different masses and thus
released to the interstellar medium on different timescales. 
We have used two different sets of measurements: 
\begin{itemize}
\item dataset~\#1: abundances from high resolution optical and NIR spectroscopy of bulge field \citep{johnson_14, rich05,rich07,rich12} and Ter5 \citep{origlia04,origlia11,origlia13} giants.
As a control-sample of halo/disc stars we used the homogeneous compilation by \citet{bensby14}.
The quoted abundances have been re-scaled to be placed on the common solar reference scale of \citet{magg_22}. Table~\ref{tab1} reports the corresponding scaling factors for the chemical elements under study.
\item dataset~\#2: abundances from APOGEE spectroscopy of 
bulge field stars and of a control-sample of halo-disc stars from the APOGEE allStarLite-dr17-synspec-rev1.fits catalog,\footnote{See the \url{https://www.sdss4.org/dr17/irspec/spectro_data/} webpage)} and of Ter5 giants from the compilation of \citet{apogeeGC}, which also include those discussed in \citet{taylor22}. 

\end{itemize}

\begin{figure}[t]
    \centering
    \includegraphics[width=1\linewidth]{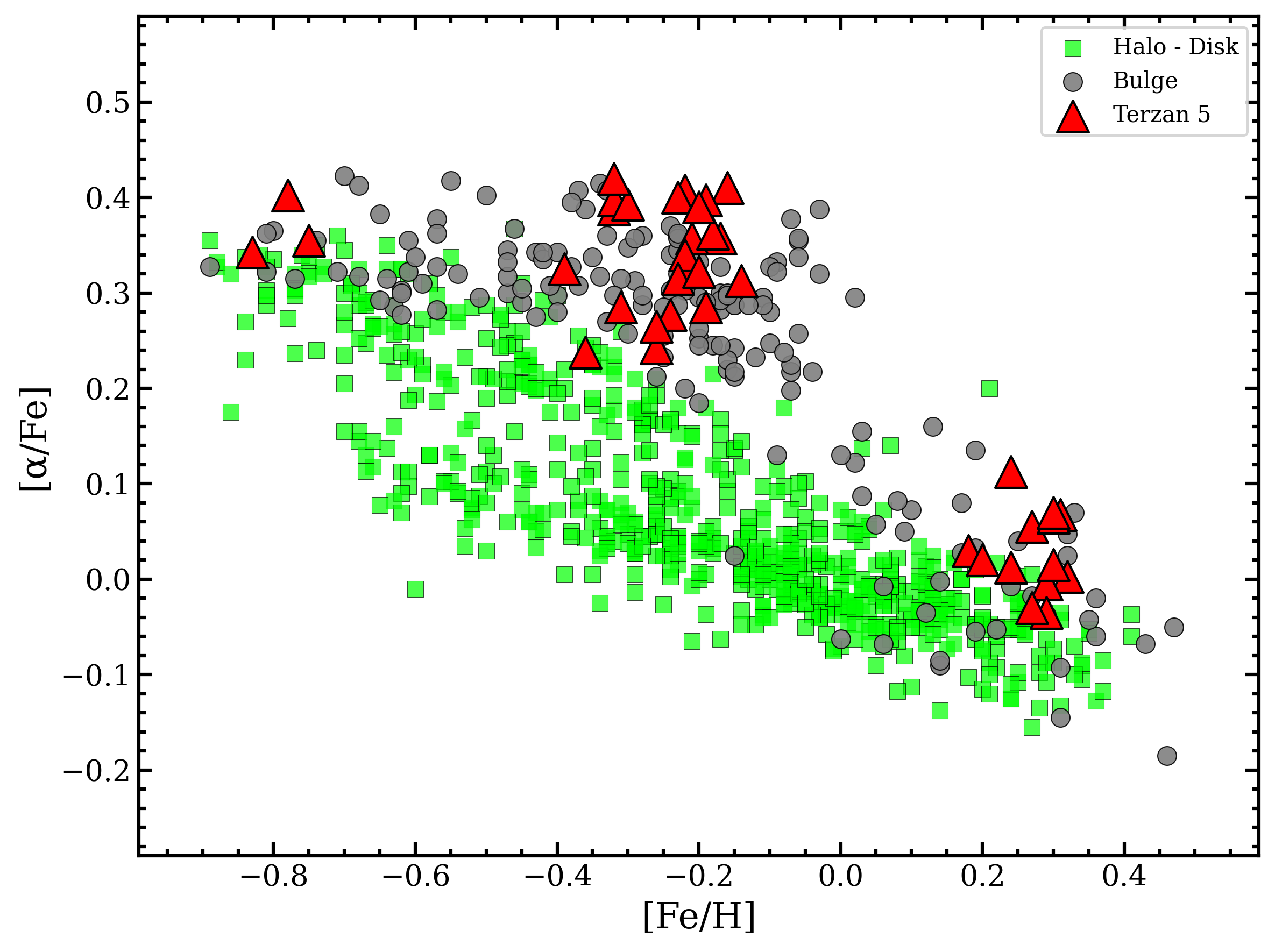}
    \caption{ Average [<Ca,Si,Mg,O>/Fe] 
    (i.e. [$\alpha$/Fe]) trends with [Fe/H] of the bulge (gray dots), halo-disc (green squares) and Ter5 (red triangles) stars from dataset \#1 (see Sect.~\ref{link}).}
    \label{alpha}
\end{figure}

\begin{figure*}
    \centering
    \includegraphics[width=0.85\linewidth]{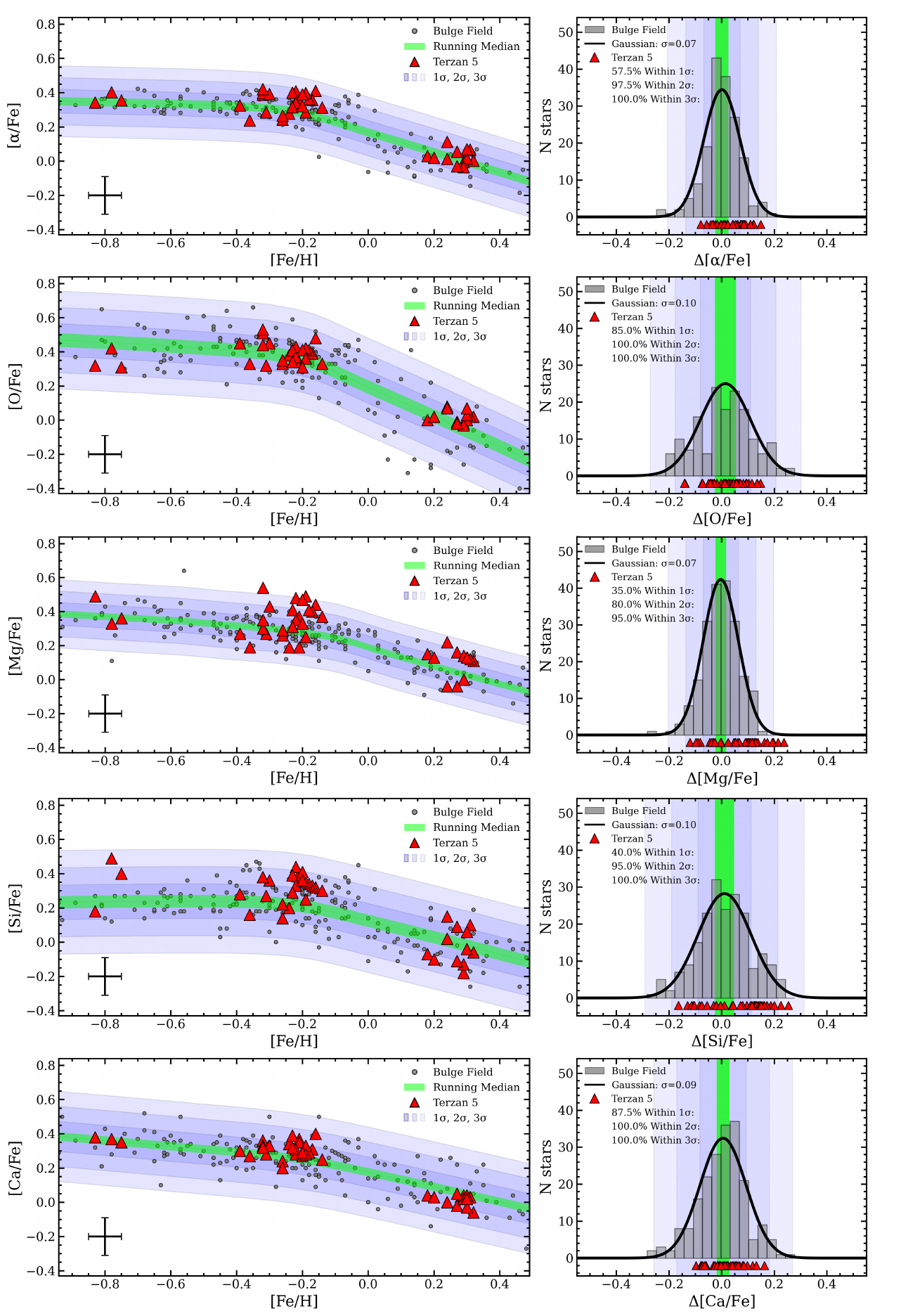}
    \caption{Chemical abundance ratio distributions of the bulge field  and Ter5 stars from dataset~\#1 (see Sect.~\ref{set1}). Left panels: [Si/Fe], [Ca/Fe], [Mg/Fe], [O/Fe] and the average [<Ca,Si,Mg,O>/Fe] 
    (i.e. [$\alpha$/Fe]) trends with [Fe/H] of the bulge field (gray dots) and Ter5 (red triangles) stars. The running LOWESS median curves at 95\% confidence (green, thick lines) and 1,2,3~$\sigma$ shaded regions are also plotted,  together with the typical measurement errorbars. Right panels:  corresponding distributions of the differences with the LOWESS median curve values.}
    \label{ratios}
\end{figure*}

\begin{figure*}
    \includegraphics[width=\textwidth]{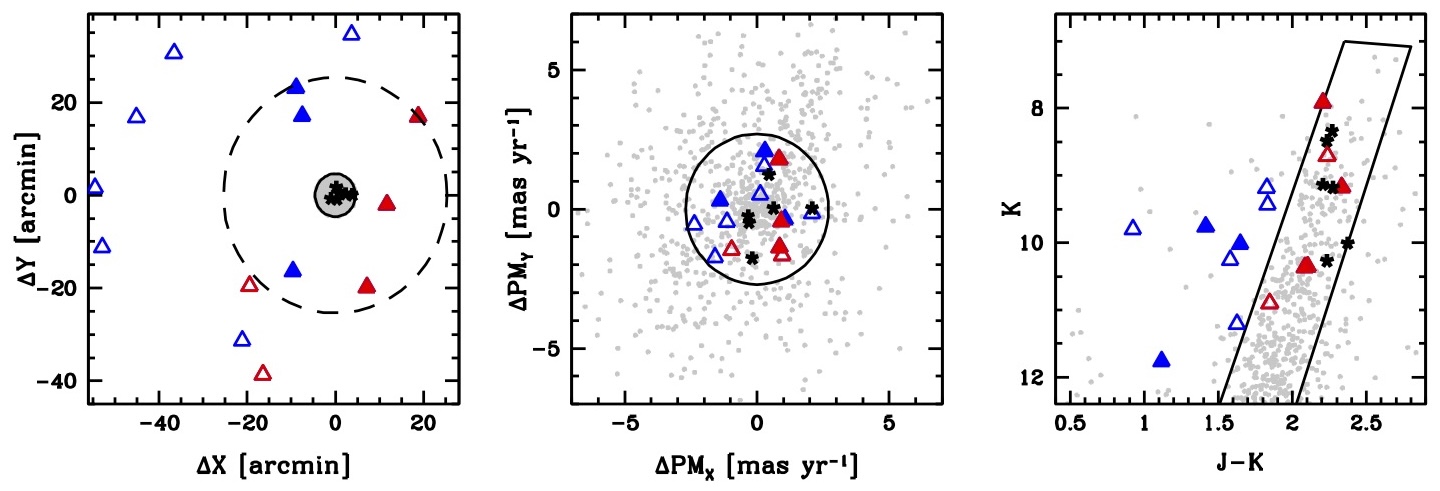}
    \caption{Diagnostic diagrams for target selection.
    Left panel: spatial map projected on sky of the Ter5 stars from the compilation by \citet{apogeeGC}, 
    referred to its center at RA=$263\rlap{.}^\circ3523333$, Dec=$-33\rlap{.}^\circ3895556$. The black, solid circle has the tidal radius of 4.6' quoted by \citet{lanzoni10}, the black dashed circle has the Jacobi radius of 25.37' quoted by \citet{taylor22}.
    Middle panel: Gaia DR3 proper motions referred to the systemic values of Ter5 quoted by \citet{vasiliev_21} for the Ter5 and the bulge field stars within the tidal radius. The black solid circle has the proper motion confidence radius of 2.7~mas~$\rm yr^{-1}$ quoted by \citet{taylor22}.   
    Right panel: corresponding 2MASS CMD for the stars in the middle panel with proper motions within the confidence radius. The black rectangle delimits the Ter5 giant branch sequences. 
    In all panels symbols are as follows: gray dots for stars within the tidal radius, black stars for APOGEE Ter5 stars within the tidal radius, filled triangles for APOGEE stars within the Jacobi radius and empty triangles for stars outside it. Blue triangles mark stars with J-K colors significantly bluer than those of Ter5 giants, hence likely foreground stars, while red triangles mark stars with J-K colors consistent with those of Ter5 giants.}
    \label{sel}
\end{figure*}

\subsection {Abundance ratio trends from dataset~\#1}  
\label{set1}

The distribution of the average [$\alpha$/Fe] abundance ratio with [Fe/H] for the Ter5 sample (40 stars) and for the bulge (about 200 stars) and halo-disc (about 700 stars) control samples from dataset \#1 is plotted in Figure~\ref{alpha}. A qualitative agreement between the Ter5 and the bulge distributions and a significant discrepancy with the halo-disc one can be noticed.

Hence, in order to evaluate on a more quantitative ground the possible chemical link between Ter5 and the bulge, 
we then modeled the [Ca/Fe], [Si/Fe], [Mg/Fe], [O/Fe] and the average 
[$\alpha$/Fe] distributions with [Fe/H] of the bulge stars
with a Locally Weighted Scatterplot Smoothing \footnote{\url{https://www.statsmodels.org/devel/generated/statsmodels.nonparametric. smoothers lowess.lowess.html}} (LOWESS) function, which is a non-parametric regression technique to obtain an optimal median curve.
This methodology involves the selection of a suitable window of neighbouring points around each point of the dataset, assigning weights (according to their distance from the selected one) and fitting a local regression function within that window. 

In order to compute the dispersion of the observed distributions around their LOWESS median function, 
we used both a boostrapping technique and a maximum-likelihood method assuming a Gaussian distribution. 
Since results are very similar with both methods, with minor (if any) asymmetries above and below the LOWESS curve and across the sampled [Fe/H] range, we adopted a single Gaussian dispersion value for each analyzed distribution. We found 1$\sigma$ values in the 0.05-0.1 dex range.
It is worth noticing that these Gaussian dispersions are also fully consistent (within 0.01 dex) with the corresponding standard deviations of the observed distributions. 
We then compared the bulge distributions with the corresponding ones of Ter5. 
By comparing the abundance distributions of bulge stars from \citet{johnson_14} and \citet{rich05,rich07,rich12} optical and NIR datasets, respectively, we estimated that possible systematics (differences in the median values) between the two analysis are well within 0.1 dex for the chemical elements under study. Since the Ter5 dataset from \citet{origlia04,origlia11,origlia13} used the same NIR spectroscopy and analysis of \citet{rich05,rich07,rich12} one can expect similar systematics between \citet{johnson_14} and Ter5 datasets.

Figure~\ref{ratios} shows 
the [Si/Fe], [Ca/Fe], [Mg/Fe], [O/Fe] and the average [<Ca,Si,Mg,O>/Fe] trends with [Fe/H] of the bulge and Ter5 stars from dataset~\#1 (left panels) and the corresponding distributions of the differences between the Ter5 values and the corresponding running LOWESS median curve ones (right panels).  
As can be appreciated, all the measured stars in Ter5 are located within 3$\sigma$ from the bulge distributions, the vast majority (between 80 and 100\%) within  2$\sigma$ and a significant (always exceeding 30\% and up to almost 90\%) fraction within 1$\sigma$. Such a statistics should be only marginally (if any) affected by the possible systematics between the bulge and Ter5 datasets mentioned above, since such systematics are always smaller than the 1~$\sigma$ values of the distributions.

We finally provide an estimate of the probability that Ter5 stars belong to either the bulge or the halo-disc.  
To this purpose, we employed a Kernel Density Estimator (KDE)  to model the density probability of the control-samples (either the bulge or the halo-disc) in four dimensions, i.e. [O/Fe], [Mg/Fe], [Si/Fe], [Ca/Fe] and in the same [Fe/H] range covered by Ter5 stars( i.e. between approximately -1.0 and +0.5 dex) . Then, we estimated the probability that the target sample belongs to each control-sample using a likelihood function derived from the KDEs and taking also into account the errors of the abundance measurements. These probabilities have been then normalized as follows.

P$_{\rm norm}$ [bulge] = P[bulge]/(P[bulge] + P[halo-disc])

P$_{\rm norm}$ [halo-disc] = P[halo-disc]/(P[bulge] + P[halo-disc])

In order to verify the ground-truth of the adopted procedure, we used a bootstrap technique with 1000 random extractions of a number of stars equal to the size of the Ter5 sample from one control-sample and we evaluate the probability that they belong to the other control sample. 
For dataset \#1, we thus extracted 1000 times 40 stars from the bulge and halo-disk control samples and we obtained a similar probability of less than 2\% that they belong to the other sample. 
The inferred probability that the Ter5 stars globally belong to the bulge sample is  100\% (median value) or 99.7\% (weighted mean with a standard deviation of 5\%).
We also computed the values separately for the sub-solar (27 stars) and super-solar (13 stars) components of Ter5, finding probabilities of 100\%  and  94\% that they belong to the corresponding bulge sub-solar and super-solar sub-samples, respectively.

\begin{figure*}
    \centering
    \includegraphics[width=0.85\linewidth]{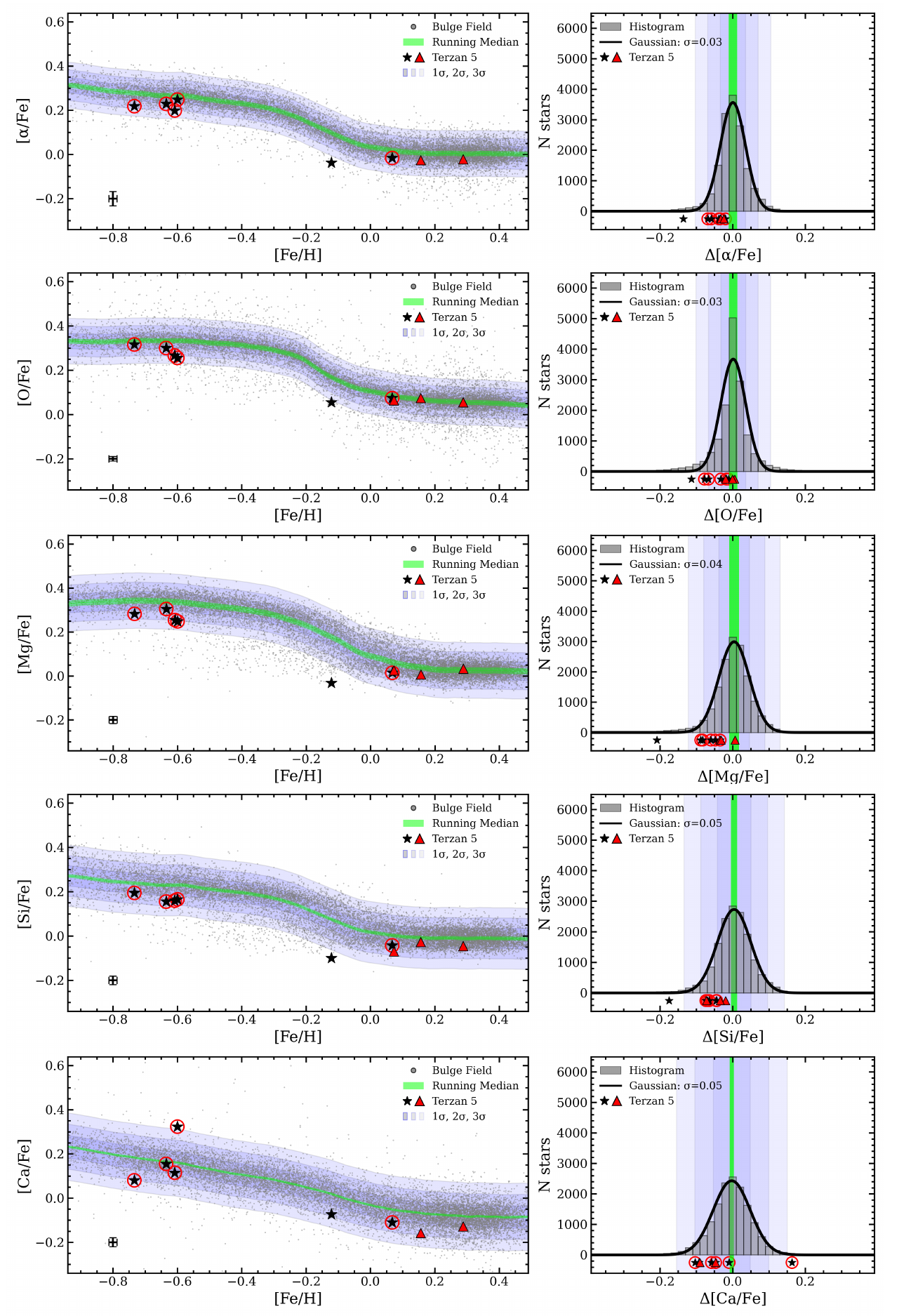}
    \caption{APOGEE DR17 chemical abundance ratio distributions of the bulge field and Ter5 stars from dataset~\#2 (see Sect.\ref{set2}). Left panels: [Si/Fe], [Ca/Fe], [Mg/Fe], [O/Fe] and the average [$\alpha$/Fe]) trends with [Fe/H] of the bulge field (gray dots) and Ter5 (black stars and red triangles) stars. Black stars refer to Ter5 members within the tidal radius of 4.6' quoted by \citet{lanzoni10}, while red triangles mark stars outside 4.6' and within the Jacobi radius of 25.4' quoted by \citet{taylor22}. The black stars circled in red are in common with \citet{taylor22}. 
    The running LOWESS median curves at 95\% confidence (green, thick lines) and 1,2,3~$\sigma$ shaded regions are also plotted, together with the typical measurement errorbars. Right panels: corresponding distributions of the differences with the LOWESS median curve ones.}
    \label{ratios_apo}
\end{figure*}

\subsection{Abundance ratio trends from dataset~\#2}
\label{set2}

\begin{table}
\tiny
\renewcommand{\arraystretch}{1.15}
\setlength{\tabcolsep}{5.5pt}
\caption{Ter5 member stars from the APOGEE sample.}
\label{tab2}
\begin{tabular}{|l|c|c|c|c|c|}
\hline\hline
APOGEE ID  &  [Fe/H]  &  [O/Fe]  &  [Mg/Fe]  &  [Si/Fe]  &  [Ca/Fe]  \\
\hline
2M17480088-2447295 & -0.60 & 0.26 &  0.25 &  0.17 &  0.32 \\
2M17480576-2445000 & -0.63 & 0.30 &  0.30 &  0.16 &  0.16 \\
2M17480668-2447374 & -0.61 & 0.27 &  0.25 &  0.16 &  0.11 \\
2M17480857-2446033 & -0.73 & 0.32 &  0.28 &  0.19 &  0.08 \\
2M17481414-2446299 &  0.07 & 0.07 &  0.01 & -0.04 & -0.11 \\
2M17482019-2446400$^*$ & -0.12 & 0.06 & -0.03 & -0.01 & -0.07 \\
2M17483400-2506387 &  0.07 & 0.06 &  0.02 & -0.07 &   --  \\
2M17485203-2448493 &  0.16 & 0.07 &  0.01 & -0.03 & -0.16 \\
2M17492070-2429466 &  0.29 & 0.05 &  0.03 & -0.05 & -0.13 \\
\hline\hline
\end{tabular}
$^*$ This star has a COLORTE\_WARN, indicating some discrepancy between the spectroscopic and color temperatures.
\end{table}

The selection of the APOGEE  bulge star sample by \citet{taylor22} was distance-based, assuming a spherical bulge with a radius of 4~kpc. Although we verified that the resulting distributions at least qualitatively are very similar, we preferred to use somewhat different criteria, namely
$|l|<20^{o}$, $|b|<10^{o}$, where \textit{l} and \textit{b} are the Galactic longitude and latitude, respectively, to properly sample the barred/peanut-shaped bulge; $<$0.25 kpc$^{-1}$ positive parallaxes to clean the sample from the most obvious foreground stars; T$_{eff}$$<$5000~K and log(g)$<$2.8 dex to sample the giant stars down to the red clump; ASPCAPFLAG = 0 to include only the stars with good spectra and reliable abundances. 

The above selection provided us with a sample counting almost 16,000 giant stars in the direction of the Galactic bulge. We did not apply any further  selection in parallax and/or in parallax error since it is very likely that bulge stars, especially those in the inner region, have (if any) very uncertain Gaia measurements of these quantities, due to the high reddening (thus implying  faint G-magnitudes) and crowding in that direction.
We also selected a similar control sample of about 17,000 halo/disc giant stars within 1.3 kpc from the Sun.

Concerning Ter5, we used the recent compilation by \citet{apogeeGC} which counts 20 stars toward this stellar system with FE\_H\_FLAG=0 
(i.e. with reliable iron abundances) and GAIA EDR3 proper motions within the confidence radius of 2.7~mas~$\rm yr^{-1}$ quoted by \citet{taylor22}. 
These stars have heliocentric RVs between -116 and -36.4 km~s$^{-1}$, generously consistent with the Ter5 systemic value of about -82~km~s$^{-1}$ and a dispersion of $\le$10~km~s$^{-1}$ at radial distances $\ge$2 arcmin \citep[see][and references therein]{baumgardt_18}\footnote{{\it Fundamental parameters of Galactic globular clusters}, \url{https://people.smp.uq.edu.au/HolgerBaumgardt/globular/}}.
This sample also includes the seven stars analyzed by \citet{taylor22}.

Figure~\ref{sel} shows the distribution of the stars toward Ter5 projected on sky, their Gaia DR3 proper motions and the 2MASS color-magnitude (CMD) diagram.
Among the 20 stars in the Ter5 compilation by \citet{apogeeGC} with proper motions within the confidence radius, only twelve are within the Jacobi radius of 25.37' quoted by \citet{taylor22}, and only six are within the tidal radius of 4.6' quoted by \citet{lanzoni10}, while the remaining eight stars outside the Jacobi radius are very likely field stars.

From the distribution of ~1500 radial velocities measured out to ~13' from the center (Ferraro et al. 2025, in preparation) we estimate that the probability of observing a giant star member of the system at distances larger than 5' is of a few percent only.  Nevertheless, to maintain an overall consistency with the assumptions made by \citet{taylor22}, we still considered members of Ter5 those stars from dataset~\#2 that are within the Jacobi radius adopted by \citet{taylor22}.

Location within the Jacobi radius and proper motions within the confidence radius are not sufficient to univocally identify Ter5 member stars. Indeed, among the six stars located between the tidal and the Jacobi radius, three have (J-K) colors significantly bluer (by at least half a magnitude) than the stars within the tidal radius, too blue to be explained in terms of differential reddening, hence they are likely foreground stars. Two of these, likely foreground stars, are also in the \citet{taylor22} sample.
Thus, in the following, we conservatively assumed nine candidate members of Ter5 (see Table~\ref{tab2}), the six black stars and the three red filled triangles in Fig.~\ref{sel}, i.e. the stars lying within the
Jacobi radius (dashed circle in the left panel of Fig.~\ref{sel}), the proper motion confidence radius (solid circle in the middle panel of Fig.~\ref{sel}) and the Ter5 red giant branch locus (the solid box in the right panel of Fig.~\ref{sel}).  As can be seen in Table~\ref{tab2}, among these nine candidates, one star lacks the 
measurement of the Ca abundance.

Figure~\ref{ratios_apo} shows the [O/Fe], [Mg/Fe], [Si/Fe], [Ca/Fe] and the average [<O,Mg,Si,Ca>/Fe] trends with [Fe/H] of the bulge and Ter5 stars from dataset~\#2 (left panels) and the corresponding distributions of the differences between the Ter5 values and the corresponding LOWESS median ones (right panels).  
It is interesting to note that the three most metal rich stars in the sample (red filled triangles in Fig.~\ref{ratios_apo}) are located in the outer regions of Ter5 (see middle panel of Fig.~\ref{sel}). Also, the two most metal-rich ones have RVs that are at 3-5 $\sigma $ from the systemic, thus barely consistent with a membership, given also that this super-solar subpopulation, if truly member of Ter5, should be centrally concentrated \citep[see][]{ferraro09,lanzoni10}.

Similarly to what found from  dataset \#1, the Ter5 likely member stars from the APOGEE sample (black stars and possibly red triangles in Fig.~\ref{ratios_apo}) are located well within 2$\sigma$ from the bulge distributions with the only exception of one star at [Fe/H]=-0.12 that shows some more discordant (between 2 and 5 $\sigma$) abundance ratios. This star has ASPCAPFLAGS =  COLORTE\_WARN, indicating some discrepancy between the spectroscopic and color temperatures.
Interestingly, five out of eight stars of Ter5 have average [$\alpha$/Fe] values that match those of the bulge distribution within 1 $\sigma$, almost coinciding with the median curve.

Although this Ter5 sample of stars is barely representative of such a complex stellar system, for sake of 
consistency with what done for dataset \#1, we also computed the probability that the 8 Ter5 stars, with measurements for all the four $\alpha$-elements, belong to the bulge sample, finding values of  93\% (median) or 98\% (weighted mean with a standard deviation of 9\%).

Thus, our analysis demonstrated that also the Ter5 APOGEE sample, although quite small in size, turns out to have a chemistry fully compatible with the one of the Galactic bulge at  2$\sigma$ level, and somewhat in contrast with the conclusion by \citet{taylor22} that their abundance patterns differ to a reasonable degree of statistical significance.

\subsubsection{The \citet{taylor22} analysis}
\label{T22}

\citep{taylor22} questioned the chemical link between Ter5 and the bulge based on the analysis of 5-7 giant stars, candidate members of Ter5, observed with APOGEE.  They adopted a random sampling technique and concluded  that the median abundance ratios, as computed from the corresponding values measured in those seven stars are statistically different from the median curves drawn by their bulge counterparts. However, some considerations about the suitability of the applied statistical analysis are worth to be done.

The small 95\% confidence intervals of the running median curves derived by \citet[][see their Figs. 3 and 4 and discussed in their Sect.~3.1]{taylor22} are the errors of these median curves, while the spreads ($\sigma$) of the distributions are clearly significantly larger.
The two quantities should not be confused. 
Indeed, for the bulge sample, we also computed the 95\% confidence errors of the running LOWESS median curves (corresponding to the thickness of the solid green lines in Fig.~\ref{ratios_apo}) via a bootstrapping technique, as done by \citet{taylor22}, and we consistently found small values, significantly smaller than the spreads of the distributions, implying that only a rather small fraction (i.e. as small as 8\% and at up to 27\% at most) of the bulge stars are located within those limits, the large majority of them being much more spread around the median curves.
Hence, the similarity between Ter5 and the bulge cannot be assessed on the basis of their coincidence with the median curves, otherwise one should 
also illogically conclude that the large majority of the bulge stars are not "bulge". It should be evaluated according to the spreads of the distributions, as reported in Sects.\ref{set1} and \ref{set2}.

\citet{taylor22} also performed a statistical test based on the comparison of the median values of the [X/Fe] abundance ratios, as computed for the small number of Ter5 stars and for an equivalent small number of stars randomly extracted (using a bootstrapping technique) from the bulge distribution. 
By construction, this test seems poorly adequate to probe the Ter5 chemistry and its consistency with the one of the bulge for at least the following main reasons.
{\it i})~an unique median value of [X/Fe], regardless of star metallicity (and likely age), cannot be representative of the complex multi-age, multi-iron and multi-[$\alpha$/Fe] stellar subpopulations of Ter5 (and also of the bulge); {\it ii})~a median value is also poorly representative of the global properties of distributions that are quite scattered around that value, as it is the case for the bulge and likely also for Ter5; {\it iii})~the size of the Ter5 sample is definitely too small (five stars in total, given that the other two in their sample are likely foreground stars because of their too blue colors,  see Sect. 3.2) for any statistical variable to be robustly and safely predictive;  {\it iv})~this test can only provide information (if any) on the similarity of Ter5 with the bulge median curves, not with the bulge distributions. 

Hence the \citet{taylor22} conclusion that the chemistry of Ter5 is not consistent with that of the Galactic bulge "at statistically significant levels" does not seem appropriate.

\section{Discussion and conclusions}

The results of the statistical tests performed on the Ter5 chemical distributions (see Sect.3.1 and 3.2) are consistently pointing to a chemical link between Ter5 and the bulge and to a bulge in-situ formation and evolution, in full agreement with the kinematic and evolutionary findings (see Sect.~\ref{properties}).
However, in this respect, it is worth mentioning that Ter5 and the bulge chemical patterns have not to be strictly identical to probe an in-situ formation and evolution of Ter5. 
Indeed, the currently available  bulge field distributions are traced by stars spread over several degree wide regions, often mostly in the bulge outskirt, and much larger than the sub-degree extension (projected on sky) of bulge stellar sub-systems like Ter5, thus providing chemical information averaged over large scales, only, without properly accounting for possible gradients/inhomogeneities on the smaller scales. In addition, Ter5 is much denser than the field and potentially only a fragment of the early bulge that may have evolved and self-enriched as an independent sub-system with its own history, not necessarily identical to the one traced by an "average" bulge. Hence as a matter of fact, some  deviations from the average bulge trends of specific abundance patterns in Ter5 stars should not be a surprise, thus possibly unveiling some peculiarities in the Ter5 evolution.

In summary, all the observational facts collected so far on Ter5 have provided robust evolutionary, kinematic and chemical evidences of a bulge in-situ formation and evolution for this stellar system. Also, as discussed in Sect.~\ref{properties} 
its multi-age, multi-iron and multi [$\alpha$/Fe] stellar populations cannot be explained in terms of a single merging/accretion event of two simple sub-structures, like two globulars or a globular and a giant molecular cloud.
In order to better account for its complexity, we thus made the working hypothesis that Ter5 
might be the relic of a pristine, more massive sub-structure of the early bulge that survived total disruption, and evolved independently, likely experiencing different episodes of star formation and self-enrichment. 
This scenario is further supported by the chemical evolution model recently proposed by \citet{romano23}, who have shown that a $4\times10^7M_{\odot}$ clump (i.e almost 3 orders of magnitude less massive than the Bulge \citep[see, e.g.][]{valenti16}) can fully reproduce the Ter5 chemical abundance patterns, without the need of invoking a massive (i.e. of nearly the mass of the bulge \citealt{taylor22}) and dark-matter dominated proto-system for the occurrence of self-enrichment, as generally believed.

The possible sequence of star formation events able to reproduce the Ter5 evolutionary  properties in a self-enriching scenario
has been recently reconstructed \citep{crociati24}. An optimal match to the differential reddening and proper-motion cleaned observed CMD
can be obtained by means of synthetic CMDs that assume a complex star formation history, foreseeing an old, main episode occurred between 12 and 13 Gyr ago that generated $\sim 70\%$ of the current stellar mass (this estimate is fully in agreement with the one reported by \citep{ferraro09,lanzoni10}, followed by an extended lower-rate star formation activity that lasted several Gyrs with one, possibly two, main additional bursts, the most recent one occurred approximately 4 Gyr ago (in agreement with the age measured by \citet{ferraro16}.

The proposed scenario also offers a natural explanation of the surprisingly large population of MSPs of Ter5. In fact, the massive Ter5 progenitor could have been able to retain the ejecta and remnants of the large number (10$^4$) of type II SN explosions  that characterized its chemical evolution \citep{romano23}, thus producing the required large populations of neutron stars (NSs) since its early stages of evolution. The exceptionally large collision rate of Ter5 \citet{verbunt87,lanzoni10} favoured the recycling of (at least) a fraction of this huge population of NSs, thus generating the large population of MSPs observed today.

At the moment, the described scenario seems the most plausible since it takes into account all the observational facts currently known on Ter5.
Additional information on the detailed chemistry of the subpopulations of Ter5 are expected in the near future. In fact, high resolution spectroscopic observations for an additional large sample of giant stars of Ter5 are going to be completed  at the ESO-VLT with the spectrograph CRIRES+, in the context of the "The Bulge Cluster Origin" (BulCO) survey \citep{ferraro25}, a Large Programme specifically aimed at measuring accurate chemical abundances for the most important elements in a statistically significant sample of massive clusters of the Galactic bulge, thus providing the chemical signatures for unveiling their origin and evolution.

\begin{acknowledgements}

LO and CF acknowledge the financial support by INAF within the VLT-MOONS project. DM acknowledges financial support from the European Union – NextGenerationEU RRF M4C2 1.1  n: 2022HY2NSX. "CHRONOS: adjusting the clock(s) to unveil the CHRONO-chemo-dynamical Structure of the Galaxy” (PI: S. Cassisi). This work is part of the project Cosmic-Lab at the Physics and Astronomy Department “A. Righi” of the Bologna University (\url{http:// www.cosmic-lab.eu/ Cosmic-Lab/Home.html}).
\end{acknowledgements}

\bibliographystyle{aa}
\bibliography{mybib} 

\end{document}